\documentclass{iau}
\usepackage{graphicx}
\usepackage{wrapfig}
\usepackage[colorlinks=true,linkcolor=blue]{hyperref}
\usepackage{xspace}

\newcommand{\figref}[1]{Fig.~\ref{#1}}

\newcommand{\LCDM}{\normalfont{$\Lambda$CDM}}
\newcommand{\MV}{\ensuremath{M_{\rm V}}}

\newcommand{\vpeak}{\ensuremath{v_\mathrm{peak}}}

\newcommand{\unit}[1]{\ensuremath{\mathrm{\,#1}}\xspace}
\newcommand{\keV}[1][]{\def\tst{#1}\ifx\tst\empty{\unit{keV}}\else{\ensuremath{{#1}\, \unit{keV}}}\fi}
\newcommand{\kpc}[1][]{\def\tst{#1}\ifx\tst\empty{\unit{kpc}}\else{\ensuremath{{#1}\, \unit{kpc}}}\fi}
\newcommand{\Msun}[1][]{\def\tst{#1}\ifx\tst\empty{\unit{M_\odot}}\else{\ensuremath{{#1}\, \unit{M_\odot}}}\fi}

\title[Satellite population of the Milky Way and WDM particle mass constraints] 
{The Milky Way's total satellite population and constraining the mass of the warm dark matter particle}

\author[O. Newton et al.]   
{Oliver Newton\thanks{E-mail: oliver.j.newton@durham.ac.uk},
Marius Cautun,
Adrian Jenkins,
Carlos S. Frenk and
John C. Helly}

\affiliation{Institute for Computational Cosmology, Durham University, Durham, DH1 3LE, UK}

\pubyear{2018}
\volume{344}  
\pagerange{\pageref{firstpage}--\pageref{lastpage}}
\jname{Dwarf Galaxies: From the Deep Universe to the Present}
\editors{K. McQuinn \& S. Stierwalt, eds.}
\begin{document}

\maketitle

\begin{abstract}
The Milky Way's~(MW) satellite population is a powerful probe of warm dark matter~(WDM) models as the abundance of small substructures is very sensitive to the properties of the WDM particle. However, only a partial census of the MW's complement of satellite galaxies exists because surveys of the MW's close environs are incomplete both in depth and in sky coverage. We present a new Bayesian analysis that combines the sample of satellites recently discovered by the Dark Energy Survey~(DES) with those found in the Sloan Digital Sky Survey~(SDSS) to estimate the total satellite galaxy luminosity function down to ${\MV{}{=}0}$. We find that there should be at least $124^{+40}_{-27}$~($68\%$ CL, statistical error) satellites as bright or brighter than ${\MV{}{=}0}$ within \kpc[300] of the Sun, with only a weak dependence on MW halo mass. When it comes online the Large Synoptic Survey Telescope should detect approximately half of this population. We also show that WDM models infer the same number of satellites as in \LCDM{}, which will allow us to rule out those models that produce insufficient substructure to be viable.
\end{abstract}

\firstsection
\section{Introduction}
\label{sec:Introduction}

The abundances of dark matter~(DM) haloes and their substructures are key cosmological probes as they relate directly to the primordial power spectrum (e.g. \cite[Peebles 1982]{peebles_large-scale_1982}). In particular, models such as warm dark matter~(WDM, \cite[Avila-Reese et al. 2001; Bode et al. 2001]{avila-reese_formation_2001, bode_halo_2001})
predict a cut-off in the matter power spectrum on dwarf galaxy scales that would suppress the formation of small galaxies (\cite[Bode et al. 2001; Polisensky \& Ricotti 2011; Lovell et al. 2012; Schewtschenko et al. 2015]{bode_halo_2001,polisensky_constraints_2011,lovell_haloes_2012,schewtschenko_dark_2015}). The abundance of the faintest galaxies can thus, in principle, reveal or rule out the presence of a power spectrum cut-off.

The satellite galaxies of the Milky Way (MW), which probe the faintest end of galaxy formation, offer the best environment to constrain some properties of the DM.
However, the current census of ${\sim}50$ MW satellite galaxies is highly incomplete. The most recent wide, deep surveys---such as the Sloan Digital Sky Survey (SDSS, \cite[Alam et al. 2015]{alam_eleventh_2015}) and Dark Energy Survey (DES, \cite[Bechtol et al. 2015; Drlica-Wagner et al. 2015]{bechtol_eight_2015,drlica-wagner_eight_2015})---do not cover the entire sky, and are subject to flux limits that depend on the surface brightness of and distance to the satellites. Previous estimates of the total population, made prior to the DES and based only on SDSS data, suggested that there could be at least $3\hbox{--}5$ times more still to be discovered (\cite[Koposov et al. 2008; Tollerud et al. 2008; Hargis et al. 2008]{koposov_luminosity_2008,tollerud_hundreds_2008,hargis_too_2014}).

We improve upon these estimates in three major ways. First, for the first time we use the combined SDSS and DES data, which cover an area equivalent to nearly half of the sky. Secondly, we properly account for stochastic effects---which we find to be the leading cause of uncertainty---by introducing a new Bayesian approach for estimating the total satellite luminosity function. Finally, to characterize uncertainties arising from host-to-host variation we make use of five high-resolution simulated host haloes taken from the Aquarius project (\cite[Springel et al. 2008]{springel_aquarius_2008}).

We compare our estimate of the total MW satellite count with WDM predictions to obtain robust constraints on the DM particle mass. We do so by requiring that WDM models produce at least enough substructures to match the observed Galactic satellite count (see e.g. \cite[Macci{\`o} \& Fontanot 2010; Lovell et al. 2014; Kennedy et al.2014; Schneider 2016; Bose et al. 2017; Lovell et al. 2017]{maccio_how_2010,lovell_properties_2014,kennedy_constraining_2014,schneider_astrophysical_2016,bose_substructure_2017,lovell_properties_2017}).

This contribution contains excerpts from \cite[Newton et al. (2018; hereafter N18]{newton_total_2018}), supplemented by some results of follow-on work.

\section{Methodology}
\label{sec:Methods}
\setlength{\intextsep}{0pt}
\begin{wrapfigure}{r}{0.5\textwidth}
    \vspace{-5pt}
    \begingroup
    \centering
	\includegraphics[width=0.45\textwidth]{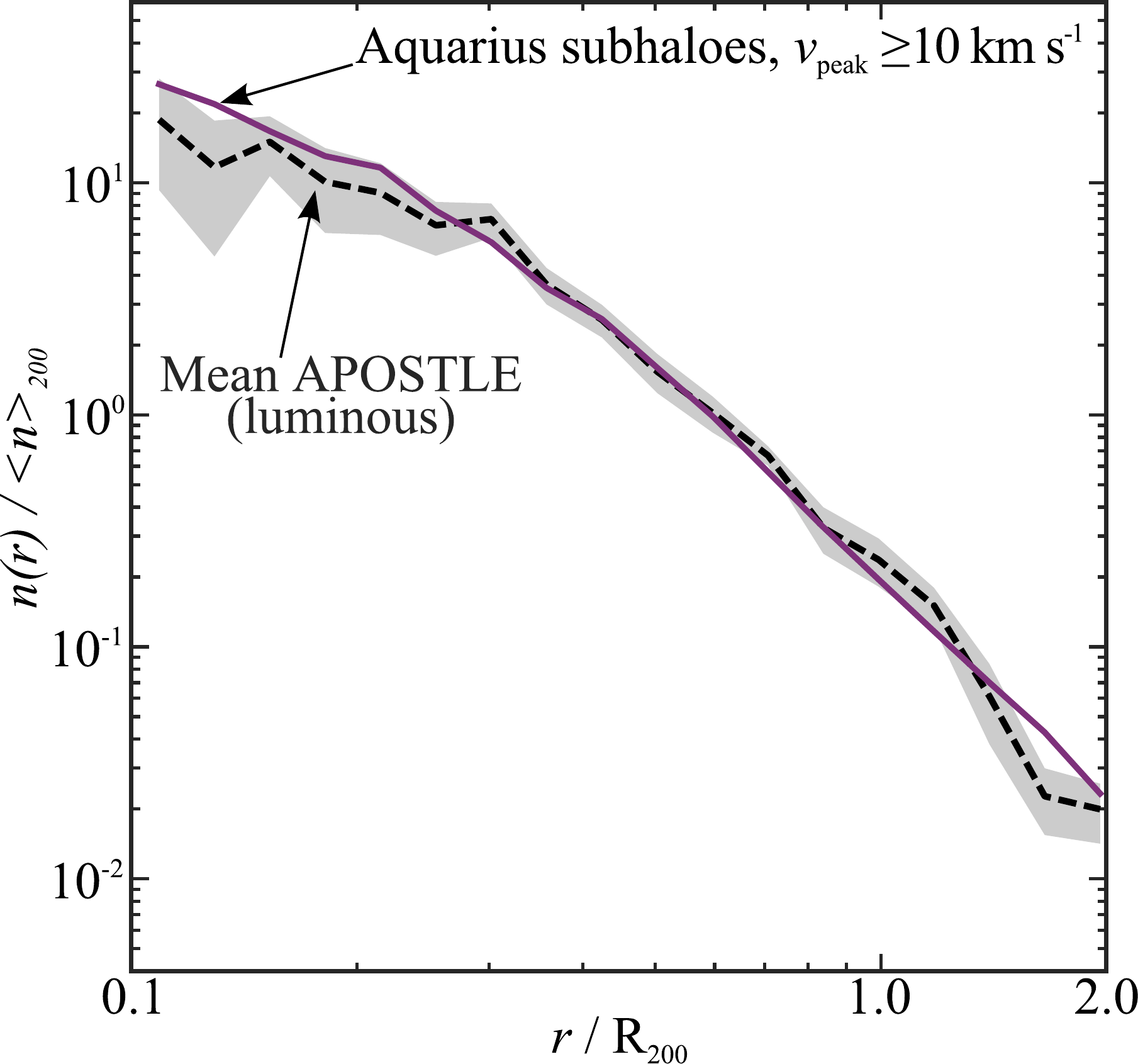}
	\caption{The radial number density of Aquarius subhaloes normalized to
          the mean density within $R_{200}$, the radius enclosing an average
          density equal to $200{\times}\rho_{\rm crit}$. The solid line shows an
          average over five Aquarius haloes. The dashed line and corresponding
          shaded region ($68\%$ scatter) is averaged over eight haloes from
          the APOSTLE high-resolution hydrodynamic simulations.}
	\label{fig:radial_density}
	\endgroup
\end{wrapfigure}

To estimate the total (all-sky) satellite galaxy population from a given survey we require two key ingredients: (i)~a prior for the radial distribution of satellites; and (ii)~our new Bayesian approach to infer the total number of satellites.

We assume that the radial number density of subhaloes in simulated MW analogues, selected by \vpeak{}---the highest maximum circular velocity attained in the subhalo's history---is the same as that of the true satellite population. The distributions of luminous satellites from state-of-the-art hydrodynamic simulations of MW analogues (\cite[Fattahi et al. 2016; Sawala et al. 2016]{fattahi_apostle_2016,sawala_apostle_2016}) and of the observed satellite population (corrected for survey radial incompleteness) support this assumption (see \figref{fig:radial_density} and \cite[N18]{newton_total_2018}, Fig.~$3$).

We construct mock SDSS and DES survey volumes consisting of two conical regions with the same sky area and spatial configuration as the corresponding real surveys. The radial extent of each mock volume is dependent on the brightness of the satellites. For each observed satellite in the combined SDSS+DES sample we calculate the maximum radius within which that satellite could have been observed in each survey and embed the resulting mock survey volumes inside each of the five Aquarius haloes. In each case, we determine the number of allowed substructures within \kpc[300] of the mock observer such that there is only one satellite within the survey footprint. We repeat the procedure for all observed satellites and for multiple pointings of the mock surveys (for more details see \cite[N18]{newton_total_2018}, section~$3$).

\section{Key Results}
\label{sec:Results}
\begin{figure}
    \centering
	\includegraphics[width=0.85\textwidth]{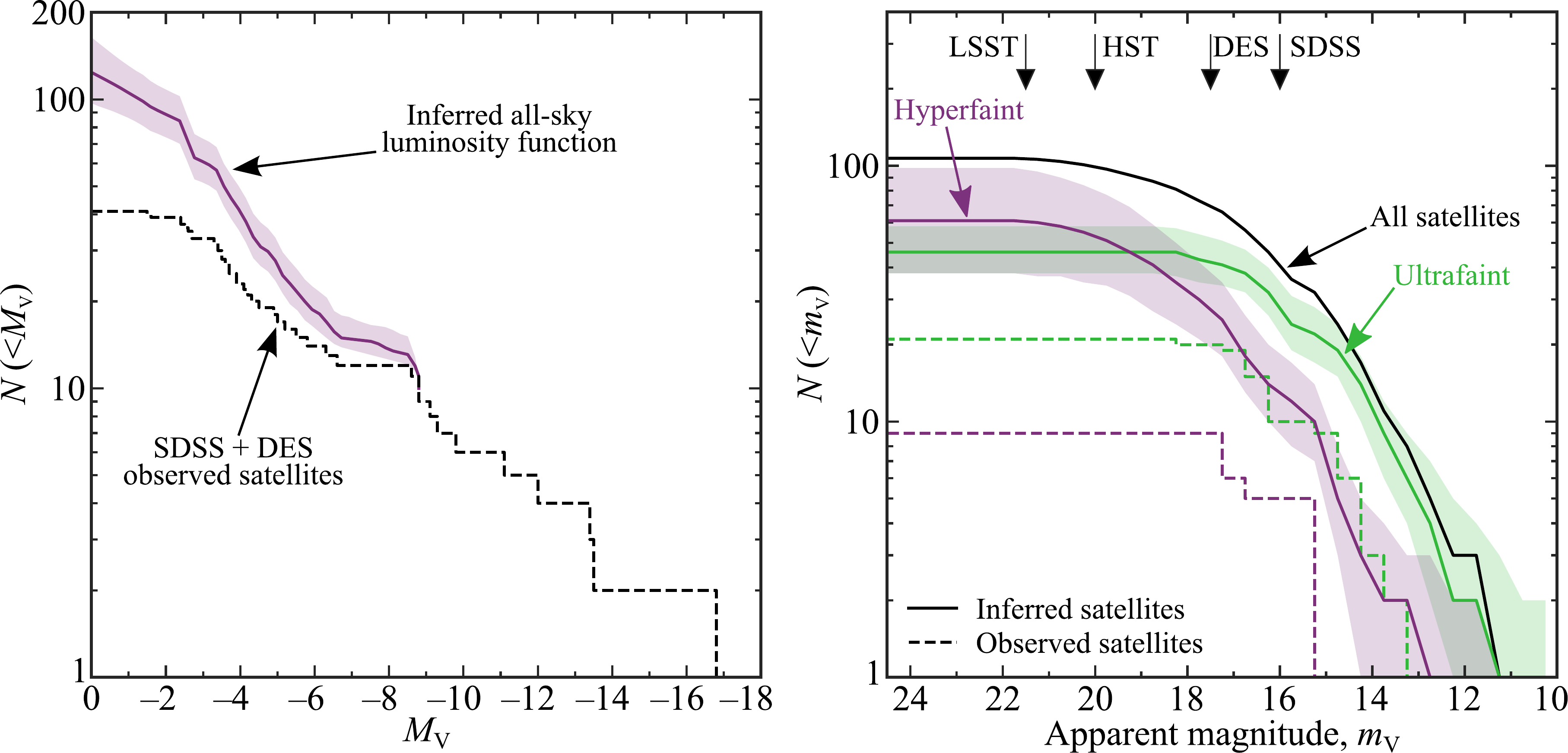}
	\vspace{-5pt}
	\caption{\textit{Left:} The total luminosity function of dwarf galaxies
	         within \kpc[300] of the Sun, obtained from combining the
	         SDSS and DES data. The solid line and the shaded region show
	         the median estimate and its $68\%$ uncertainty,
	         respectively. The luminosity function of all observed
	         satellites within the SDSS and DES footprints inside
	         \kpc[300] is indicated by the dashed line.
	         \textit{Right:} The same as the left panel but as a function
	         of \textit{apparent} $V-$band magnitude,
             $m_{\rm V}$. The satellites are split into ultra- and
             hyperfaint populations with absolute magnitude in the
             range ${-8 < \MV{} \leq -3}$ and ${-3 < \MV{} \leq 0}$,
             respectively. The solid and dashed lines and shaded regions
             have the same definitions as before. The sum of the median
             predictions of both populations is also provided (black line).
             The vertical arrows indicate the faintest satellites that can
             be detected in four past and future surveys.}
	\label{fig:satLFs}
	\vspace{-7pt}
\end{figure}

We obtain our best estimate of the total (all-sky) satellite luminosity function by combining data from the SDSS and DES (\figref{fig:satLFs}, left panel).
This depends only weakly on halo mass, as the inferred satellite count depends not on the \textit{total} number of subhaloes but on the shape of their normalized radial profile (see \cite[N18]{newton_total_2018}, section~$4.4$).

\setlength{\intextsep}{0pt}
\begin{wrapfigure}{r}{0.5\textwidth}
	\includegraphics[width=0.45\textwidth]{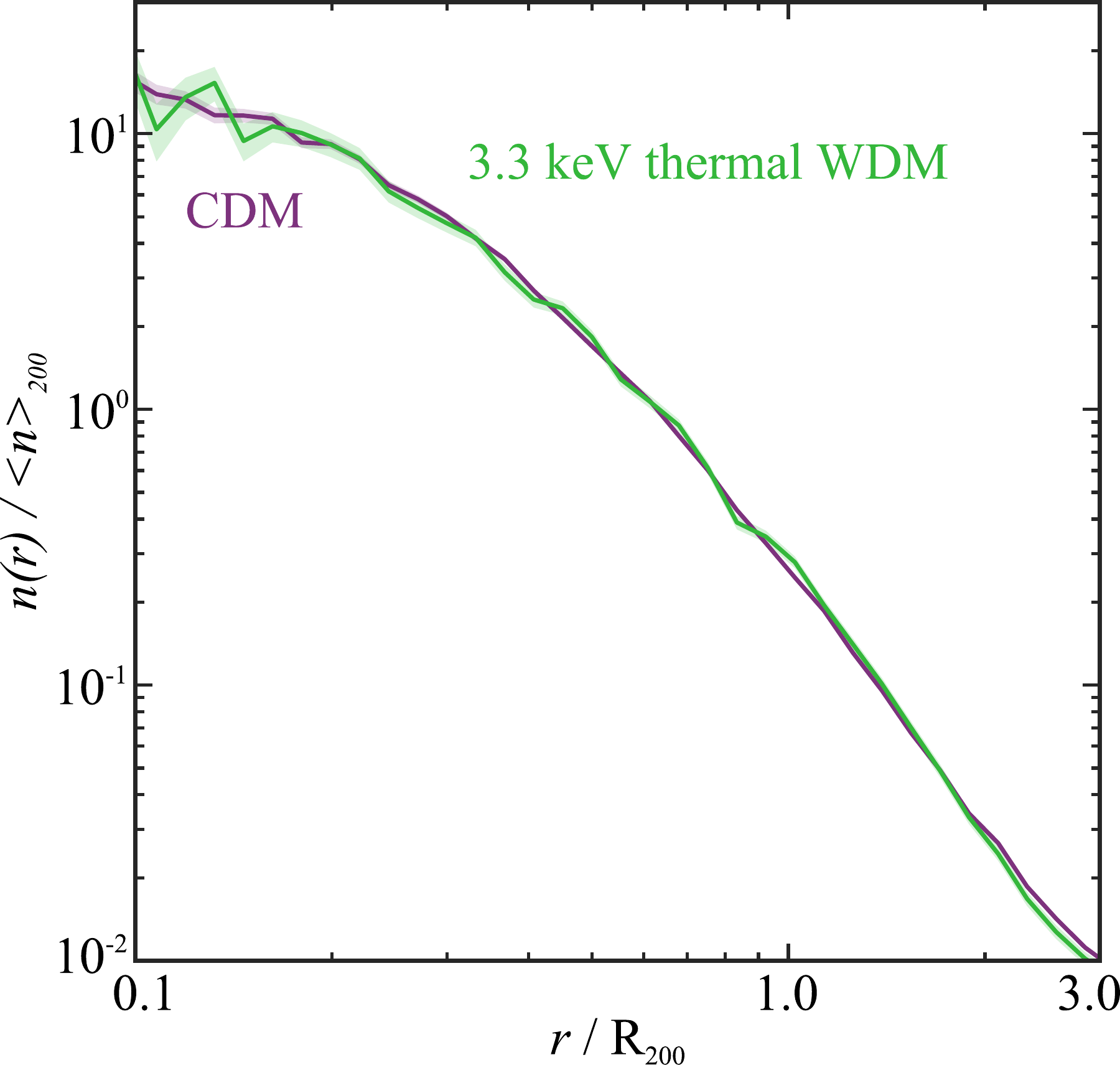}
	\caption{The radial number density of stacked COCO subhaloes normalized to the mean density within $R_{200}$. The solid lines show averages over haloes taken from the COCO-COLD~(CDM) and COCO-WARM~(\keV[3.3]) simulations. The corresponding shaded regions represent the $68\%$ scatter.}
	\label{fig:warm_radial_density}
\end{wrapfigure}

We also consider the prospects for discovery of the inferred satellite population in future surveys of the MW. To estimate apparent magnitudes, we sample the luminosity function given in the left panel of \figref{fig:satLFs} and assign absolute $V-$band magnitudes to subhaloes in the Aquarius haloes. We compute the apparent magnitude of the satellites using the distances of the subhaloes from the observer, for each observer position and generated luminosity function. The satellite galaxy counts as a function of apparent magnitude are shown in the right panel of \figref{fig:satLFs}, where `ultrafaint' galaxies have absolute magnitudes in the range ${-8 < \MV{} \leq -3}$ and `hyperfaint' satellites have ${-3 < \MV{} \leq 0}$. The Large Synoptic Survey Telescope~(LSST) will image deep enough to detect all inferred satellites, excluding any obscured by the Zone of Avoidance.

The inference method is sensitive to the radial number density of subhaloes, especially at small radii. Therefore, any DM model that produces haloes with the same radial number density as in \LCDM{} will also predict the same number of satellite galaxies. In \figref{fig:warm_radial_density} we find a close similarity between the stacked radial density profiles from the
\LCDM{} and WDM versions of the COCO simulation suite (\cite[Hellwing et al. 2016; Bose et al. 2016]{hellwing_copernicus_2016,bose_copernicus_2016}). Any WDM model in which MW-like haloes do not produce at least as many DM subhaloes as the inferred population can, therefore, be ruled out with high confidence.

\section{Conclusions}
\label{sec:Conclusion}
Until deeper, more complete surveys are undertaken we must estimate the total complement of MW satellite galaxies. This can help us to understand various astrophysical phenomena---such as the effect of reionization on small haloes (\cite[Bose et al. 2018]{bose_imprint_2018})---and to constrain the properties of DM particles.

We have, for the first time, combined data from the SDSS and DES to infer the all-sky satellite galaxy luminosity function of the MW. We predict that the MW has $124^{+40}_{-27}$~($68\%$ CL) satellites brighter than
$\MV{}{=}0$ within \kpc[300] (see \figref{fig:satLFs}) and find that this is only very weakly dependent on the assumed MW halo mass (\cite[N18]{newton_total_2018}, section~$4.4$). Once complete, the LSST survey should detect approximately half of this population. These estimates represent lower limits to the total Galactic satellite complement as they do not account for very low surface brightness objects that may have been missed in observations, nor for satellites brought in by the LMC which today lie outside the DES footprint.

We demonstrate that the radial number density of subhaloes in MW-like haloes produced by WDM models is the same as that produced in CDM models. Therefore, our estimate of the total satellite count also applies to WDM haloes and can be used to constrain the allowed mass of the WDM particle.

\section*{Acknowledgements}
ON was supported by the Science and Technology Facilities Council~(STFC) through grant ST/N50404X/1 and MC, ARJ and CSF were supported by STFC grant ST/L00075X/1.


\end{document}